\documentclass[aps,prd,superscriptaddress,amsfonts,amssymb,amsmath,showpacs,11pt]{revtex4-2}
\usepackage{silence}
\usepackage{bm}
\usepackage{accents}
\usepackage{mathrsfs}
\usepackage[T1]{fontenc}
\usepackage{tabularx}
\usepackage{graphicx}
\usepackage{colortbl}
\usepackage[caption=false]{subfig}
\usepackage{booktabs}
\usepackage{dcolumn}
\usepackage{multirow}
\usepackage{hhline}
\usepackage{placeins}
\usepackage{float}
\usepackage{amsmath}
\usepackage{amssymb}
\usepackage{array}
\usepackage[british]{babel}
\usepackage[table]{xcolor}
\usepackage{orcidlink}
\usepackage{xparse}
\usepackage[bb=boondox]{mathalpha}
\usepackage{epsfig}
\usepackage{caption}
\usepackage{subcaption}

\renewcommand{\arraystretch}{1.5}

\usepackage{hyperref}

\allowdisplaybreaks[1]
\renewcommand{\arraystretch}{1.1}
\newcommand{\midsepremove}{\aboverulesep = 0mm \belowrulesep = 0mm}
\midsepremove
\newcommand{\midsepdefault}{\aboverulesep = 0.605mm \belowrulesep = 0.984mm}
\midsepdefault
\hypersetup{
    colorlinks=true,
    citecolor=blue,
    linkcolor=blue,
    filecolor=magenta,
    urlcolor=blue
}

\begin{document}
\title{Evolutionary Phase of Universe in $f(R,L_m,T)$ Gravity: The Dynamical System Analysis}

\author{R.R. Panchal \orcidlink{xxxx-yyyy-zzzz-tttt}}
\email{ravipanchal1712@spuvvn.edu}
\affiliation{Department of Mathematics, Sardar Patel University, Vallabh Vidyanagar-388120, India.}

\author{Divya G. Sanjava \orcidlink{xxxx-yyyy-zzzz-tttt}}
\email{divyasanjava15@gmail.com}
\affiliation{Department of Mathematics, Sardar Patel University, Vallabh Vidyanagar-388120, India.}

\author{A. H. Hasmani \orcidlink{xxxx-yyyy-zzzz-tttt}}
\email{ah\_hasmani@spuvvn.edu}
\affiliation{Department of Mathematics, Sardar Patel University, Vallabh Vidyanagar-388120, India.}

\author{B. Mishra\orcidlink{0000-0001-5527-3565}}
\email{bivu@hyderabad.bits-pilani.ac.in}
\affiliation{Department of Mathematics, Birla Institute of Technology and Science-Pilani, Hyderabad Campus, Jawahar Nagar, Kapra Mandal, Medchal District, Telangana 500078, India.}

\begin{abstract}
In this paper, the dynamical system analysis has been performed to analyze the dynamical behavior of the Universe in $f(R,L_m,T)$ gravity with a scalar field. A well motivated potential function and the linear form of the functional $f(R,L_m,T)$ have been incorporated into the Friedmann equation, and the autonomous dynamical system has been framed by introducing dimensionless variables. The stability behavior of the  critical points is obtained and analyzed based on their corresponding eigenvalues. Moreover, cosmological parameters such as the deceleration parameter and the dynamical parameters such as equation of state and density parameters are obtained using the dimensionless variables. It has been observed that the system provides critical points that describe different evolutionary phases of the Universe. 
\end{abstract}
\maketitle

	\section{\label{sec:level1}Introduction}
	
	The recent cosmological study is focusing on finding an explanation for the present late time cosmic acceleration behavior of the Universe. The cosmological observations such as supernovae type Ia (SNe Ia) \cite{1supernovae,2}, Cosmic Microwave Background Radiation (CMBR) \cite{3_2020}, Baryon Acoustic Oscillations (BAO) \cite{4_Eisenstein_2005},  Wilkinson Microwave Anisotropy Probe (WMAP) experiment \cite{5_Verde_2003} have provided compelling evidence on the late time cosmic phenomena of the Universe. This behavior is believed to be induced by the negative pressure exerted by the enigmatic form of energy, called dark energy. In the mass energy budget of the Universe, the dark energy along with dark matter approximately comprises $95\%$, whereas a mere $5\%$ attribute to baryonic matter \cite{3_2020,6_Padmanabhan_2003}.  In conventional cosmological models, late time acceleration is ascribed to a cosmological constant $\Lambda$, resulting in the $\Lambda$CDM concordance model \cite{7_Peebles_2003,8_einstein1986cosmological}. However, the interpretation of $\Lambda$ as vacuum energy density \cite{9_RevModPhys.61.1,10_Bull_2016}, resulted in a discrepancy of approximately $10^{120}$ orders of magnitude between theoretical predictions and the observed value and leads to fine-tuning problem \cite{11_Harko_Lobo_2018}. Another issue that has arisen is the significant discrepancy in the present value of Hubble parameter determined by direct and model-independent late time measurements and values deduced from early Universe predictions, which is famously known as Hubble tension problem \cite{12_Riess_2022,13_DiValentino_2021}. General Relativity (GR) has been successful in explaining several issues of the Universe, but has limitations in explaining the late time phenomena. So, modification of the GR has been inevitable. Modification can be done by modifying the underlying geometry or the matter part of the Einstein field equations \cite{15_Nojiri_2007}.
	
	The curvature of space time has been taken as the framework for GR; and hence one can do the modification in the geometric part of GR \cite{14_Clifton_2012,15_Nojiri_2007}. The first modification in curvature based gravity is $f(R)$ gravity \cite{firstfrgravity10.1093/mnras/150.1.1}, in which the Ricci scalar $R$ in the action of GR has been replaced by the function $f(R)$. The higher order term of $f(R)$ facilitates exploring more complex cosmic phenomena are discussed in \cite{16_Sotiriou_2010}. The non-minimal coupling between geometry and matter, i.e. algebraic combination of the Ricci scalar $R$ and the matter Lagrangian $L_m$ leads to the formulation of $f(R,L_m)$ gravity \cite{17_Harko_2010}. Furthermore, the trace of the energy momentum tensor $T$ exhibits a non-minimal coupling to curvature, leading to the formulation of $f(R,T)$ gravity \cite{18_PhysRevD.84.024020}. Both $f(R,L_m)$ and $f(R,T)$ gravity exhibit complex phenomena on astrophysical and cosmic scales. These frameworks address the matter Lagrangian and trace of the energy momentum tensor independently, indicate the possibility to develop a more comprehensive and unified coupling system. Haghani and Harko \cite{19_Haghani_2021} introduced the unification of $f(R,L_m)$ and $f(R,T)$ gravitational theories known as $f(R,L_m,T)$ gravity.  It should be mentioned that $f(R)$, $f(R,L_m)$, and $f(R,T)$ gravitational theories serve as the limiting cases of $f (R,L_m,T)$ gravity theory.
	
	We discuss some of the recent developments in $f(R,L_m,T)$ gravity in this paragraph. Generalized $f(R,L_m,T)$ gravity, a unified matter-geometry coupling framework that results in a non-conserved energy momentum tensor can effectively explain cosmic acceleration without invoking dark energy \cite{19_Haghani_2021}. The observationally constrained model shows late time acceleration behavior with feasible ranges of the cosmological parameters \cite{B:maurya2025cosmologicalimplicationscausalityfr}. In \cite{C:maurya2025transitdarkenergycosmological}, by taking a particular non-linear form of 
	$f(R,L_m,T)$ stability, causality, and  cosmic evolution was discussed. Through dynamical system analysis, the transition from a decelerating to an accelerating Universe has been shown with the stable attractor solutions \cite{38_kshirsagar2025cosmologicaldynamicsfrlmtmodified}.   Reconstructed models are stable under perturbations and thermodynamically consistent \cite{40_https://doi.org/10.1002/prop.202300018}. Different field equations with potentially testable observational repercussions were constructed in the Palatini formalism \cite{39_deLima2025}. Cosmological models are consistent with empirical Hubble data and effectively characterize late time accelerated expansion \cite{37_MAURYA2024101722}. The matter-geometry interaction promotes stable wormhole solutions and lessens energy condition violations \cite{G:errehymy2025rolecosmicvoidsmatter} and with appropriate coupling, traversable wormholes can be obtained that require less exotic matter \cite{H:errehymy2025possiblewormholesgeneralizedgeometrymatter}. In some circumstances, wormhole solutions might occur without the need for exotic matter \cite{33_Moraes_2024}. From an astrophysical point of view, the interior structure and complexity of compact objects, where anisotropy and electromagnetic fields are essential to stability, are greatly impacted by matter-geometry coupling \cite{35_Rehman:2026mee}. Furthermore, $f(R,L_m,T)$ gravity can alter the Chandrasekhar mass limit, permitting super-Chandrasekhar white dwarfs and provide an explanation that goes beyond GR predictions for massive stellar objects \cite{E:priyobarta2026possibleexistencesuperchandrasekhar}. In addition, it is possible to create stable neutron star configurations that are consistent with current observational data \cite{32_fortunato2024hydrostaticequilibriumconfigurationsneutron}. The neutron star study \cite{34_Mota_2024} showed that the mass–radius relationship of neutron stars is greatly impacted by modified Tolman–Oppenheimer–Volkoff equation. Stellar models that were physically acceptable, stable, and in agreement with the data were obtained in \cite{36_Zubair:2026znq}.
	
	The main challenge in modified gravity theories is to obtain an analytic solution because of the non-linear nature of its field equations. So, to overcome this and frame a cosmological model of the Universe, the dynamical system analysis approach is being used  \cite{42_Samaddar:2025kyr,43_mandal2025frgtgravity,44_NARAWADE2022101020,45_BHAGAT2026100483,46_Samaddar:2025nve,Doi1,Doi2,Doi3}. In dynamical system analysis, modified gravity field equations are converted into an autonomous system using dimensionless variables.  The objective is to identify stable critical points that may be helpful in characterizing the different phases of the evolution of the Universe. The concept of a dynamical system primarily involves locating critical points in a coupled system of first-order differential equations. Consequently, the stability criteria can be determined by computing the Jacobian matrix at key sites and examining their eigenvalues \cite{41_wiggins2003introduction}. Since the nature of the dark energy is unknown, the scalar field can be used as an efficient candidate that can provide accelerated expansion through their potential and kinetic terms.  In addition to offering scaling and attractor solutions that can naturally propel the late time accelerated expansion of the Universe, scalar fields offer a dynamical description of dark energy that permits a time dependent equation of state. A thorough examination of the stability and asymptotic behaviour of cosmological solutions is also made possible by the addition of a scalar field, which provides an easy framework for dynamical system analysis. So, we may add the scalar field to the dynamical system analysis of modified gravity models. A unified account of many cosmological stages, such as early time inflation and late time acceleration can be obtained by using this approach. Furthermore, a more accurate modeling of the cosmic evolution is made possible by taking into account other components with distinct properties. The other scalar field candidates for dark energy are k-essence \cite{47_PhysRevD.63.103510}, Chaplygin gas \cite{48_COPELAND_2006}, quintessence \cite{48_COPELAND_2006}, or phantom \cite{49_Nojiri_2006}; which can provide dynamical evolution of the universe. Therefore, adding a scalar field improves the flexibility of the model  and makes it easier to understand cosmic dynamics  \cite{50_sharma2025dynamicalsystemsanalysisfrg,51_doi:10.1142/S0219887819500233,42_Samaddar:2025kyr}. Recently in \cite{38_kshirsagar2025cosmologicaldynamicsfrlmtmodified}, authors have done the dynamical system analysis of the non linear $f(R,L_m,T)=R-\mu L_mT-\gamma$ model. In the present work, a linear form $f(R,L_m,T)=R+\alpha L_m+\beta T$ is examined in the presence of a canonical scalar field with exponential potential. The current model includes an extra scalar field sector that offers additional dynamical degrees of freedom that advances the evolution of the cosmos, whereas \cite{38_kshirsagar2025cosmologicaldynamicsfrlmtmodified} concentrates on the cosmic dynamics produced by the matter geometry connection.
	
	The paper is organized as follows: in section--\ref{sec2}, the basic mathematical formalism of $f(R,L_m,T)$ gravity has been presented, and the field equations are expressed by considering a well motivated potential function. The dynamical system analysis has been performed in section--\ref{sec3} and the critical points are analyzed at different evolutionary epoch. In section--\ref{sec4} the conclusion has been given.
	\section{Basic formalism of $f(R,L_m,T)$ gravity} \label{sec2}
	The action of $f(R,L_m,T)$ gravity \cite{19_Haghani_2021} with the canonical scalar field is given by
	\begin{align}\label{1}
		S=&\int f(R,L_m,T)\sqrt{-g}d^4x + \int (L_m+L_{\phi})\sqrt{-g}d^4x,
	\end{align}
	in which $R$, $L_m$ and $T$ respectively denote Ricci scalar, matter Lagrangian and trace of energy momentum tensor; $L_{\phi}$ be the Lagrangian for the scalar field. We shall take $c=1$ and $8\pi G=\kappa ^2$. Taking the variation of (\ref{1}) with respect to metric tensor $g^{\mu\nu}$, the modified field equations \cite{19_Haghani_2021} can be written as,
	\begin{align}\label{2}
		f_R G_{\mu\nu}=& \kappa ^2 \left(T_{\mu\nu}^{(m)}+\bar{T}^{(de)}_{\mu\nu}\right),
	\end{align}
	where $ \bar{T}^{(de)}_{\mu\nu}=T_{\mu\nu}^{(\phi)}+T_{\mu\nu}^{(de)}$  represents total dark energy contribution due to the scalar field and modified gravity terms, $f_R=\frac{\partial f}{\partial R}$ and $G_{\mu\nu}$ be the Einstein tensor. The energy momentum tensor for matter, scalar field and dark energy components are respectively represented as  $T_{\mu\nu}^{(m)}$, $T_{\mu\nu}^{(\phi)}$ and $T_{\mu\nu}^{(de)}$ and their corresponding expressions are,
	\begin{align}\label{3}
		T^{(m)}_{\mu \nu}=-\frac{2}{\sqrt{-g}}\frac{\delta(\sqrt{-g}L_m)}{\delta g_{\mu \nu}},
	\end{align}
	\begin{align}\label{4}
		T^{(\phi)}_{\mu \nu}=-\frac{2}{\sqrt{-g}}\frac{\delta(\sqrt{-g}L_{\phi})}{\delta g_{\mu \nu}}
	\end{align}
	and
	\begin{align}\label{5}
		\kappa^2 T_{\mu\nu}^{(de)}=&\frac{1}{2}g_{\mu\nu}(f-Rf_R)-(g_{\mu\nu}\square -\nabla_{\nu}\nabla_{\mu})f_R+\frac{1}{2}(f_L+2f_T)(T_{\mu\nu}-L_m g_{\mu\nu})+f_T\tau_{\mu\nu}.
	\end{align}
	
	In Eq.\eqref{5}, $\square$ denotes the d'Alembert operator  and 
	\begin{align}
		\tau _{\mu \nu}=2g^{\alpha\beta}\frac{\partial ^2 L_m}{\partial g^{\mu\nu}\partial g^{\alpha\beta}}, \quad
		f_R=\frac{\partial f(R,L_m,T)}{\partial R},\quad
		f_L=\frac{\partial f(R,L_m,T)}{\partial L_m},\quad f_T=\frac{\partial f(R,L_m,T)}{\partial T}.   \nonumber
	\end{align}
	
	In this work, the field equations of $f(R,L_m,T)$ gravity are obtained using a linear form  $f(R,L_m,T)=R+\alpha L_m+\beta T$ \cite{19_Haghani_2021}, where $\alpha$ and $\beta$ are model parameters. Using the linear form, Eq. \eqref{2} reduces to 
	\begin{align}\label{6}
		G_{\mu\nu}=& \kappa ^2 (T_{\mu\nu}^{(m)}+T_{\mu\nu}^{(\phi)}+T_{\mu\nu}^{(de)}),
	\end{align}
	where
	\begin{align}\label{7}
		\kappa^2  T_{\mu\nu}^{(de)}=\frac{1}{2}g_{\mu\nu}(\alpha L_m+\beta T)+\frac{1}{2}(\alpha+2\beta)(T_{\mu\nu}-L_m g_{\mu\nu})+f_T\tau_{\mu\nu}.
	\end{align}
	
	To frame the cosmological model of the Universe, we have considered the flat Friedmann–Lemaître –Robertson–Walker (FLRW)
	space time as,
	\begin{equation}\label{8}
		ds^2=-dt^2+a^2(t)(dx^2+dy^2+dz^2),
	\end{equation}
	in which $a(t)$ is the scale factor. The Ricci scalar for the metric is $
	R=6(\dot{H}+2H^2)$, where  $H=\frac{\dot{a}}{a}$ is  Hubble parameter. Now, we consider the dust case for the classical relativistic term. Since nature of the dark energy is unknown, the dark energy pressure and density are considered of arbitrary nature and is to be decided from the results of the dynamical system analysis. So, energy momentum tensor are considered as follows
	\begin{align}
		T_{\nu}^{\mu (m) }=diag[0, 0, 0, -\rho _m],\quad
		 T_{\nu}^{\mu (\phi)}=diag[p_{\phi}, p_{\phi}, p_{\phi}, -\rho _{\phi}],\quad
		T_{\nu}^{\mu (de) }=diag[p_{de}, p_{de}, p_{de}, -\rho_{de}]. \nonumber
	\end{align}
	
	Now, taking Lagrangian density $L_m=-\rho$ in the modified Friedmann equations \eqref{6} can be written as,
	\begin{align}
		3H^2&=\kappa ^2 (\rho_{\phi}+ \rho_m+\rho_{de}),\label{9}\\
		2\dot{H}+3H^2&=-\kappa ^2 (p_{\phi}+p_{de}),\label{10}
	\end{align}
	where  $\omega=\frac{p}{\rho}$,  $\kappa^2 \rho_{de}=\frac{1}{2}(\alpha+(1-3\omega)\beta)\rho$ and $\kappa^2 p_{de}=\frac{1}{2}(\alpha \omega+(1+5\omega)\beta)\rho$.
	
	In this model, the quantity $\rho_\phi$ reflects the energy density of the canonical scalar field that is explicitly introduced in the action through the scalar field Lagrangian $L_\phi$. $\rho$  is not brought about by a separate dark energy Lagrangian. Both $\rho_\phi$ and $\rho$ can generate negative pressure and contribute to cosmic acceleration. The terminology used in this paper is in line with previous research \cite{50_sharma2025dynamicalsystemsanalysisfrg,46_Samaddar:2025nve}. For the canonical scalar field, the energy density $(\rho_{\phi})$ and the pressure $(p_{\phi})$ \cite{50_sharma2025dynamicalsystemsanalysisfrg} are given by
	\begin{align}
		\rho_{\phi}=\frac{1}{2}\dot{\phi^2}+V(\phi),\label{11}\\
		p_{\phi}=\frac{1}{2}\dot{\phi^2}-V(\phi),\label{12}
	\end{align}
	where $V(\phi)$ is the scalar potential. We consider a particular form of the potential \cite{PhysRevD.57.4686} as follows
	\begin{equation}
		V(\phi)=V_0e^{-\lambda\phi},\label{13}
	\end{equation}
	where $\lambda, V_0>0$ are dimensionless parameters.
	We assume that there is no interaction between matter energy $\rho_m$, dark energy $\rho _{de}$ and scalar field $\rho_{\phi}$. Also, the energy momentum tensor is assumed to be divergence less, which leads to the following conservation equations
	\begin{align}
		\dot{\rho_m}+3H\rho_m&=0,\label{14}\\
		\dot{\rho_{\phi}}+3H(1+\omega_{\phi})\rho_{\phi}&=0,\label{15}\\
		\dot{\rho_{de}}+3H(1+\omega _{de})\rho_{de}&=0,\label{16}
	\end{align}
	where $p_{\phi}=\omega_{\phi}\rho_{\phi}$ and $p_{de}=\omega_{de}\rho_{de}$. Now, substituting Eq. \eqref{11} and Eq. \eqref{12} in Eq. \eqref{15}, we get Klein Gordon equations \cite{PhysRevD.57.4686, 11_Harko_Lobo_2018} governing the scalar field evolution,
	\begin{equation}
		\ddot{\phi}+3H\dot{\phi}+\frac{dV}{d\phi}=0.\label{17}
	\end{equation}
	
	In Eq. \eqref{17}, $\ddot{\phi}$ and $3H\dot{\phi}$ respectively denote the acceleration of the field and the damping effect on the expansion of the Universe. The derivative of potential $V(\phi)$ represents the force exerted by the potential. Now,  substituting Eq. \eqref{11} and Eq.\eqref{12} in Eq. \eqref{9} and Eq. \eqref{10}, one can obtain
	\begin{align}
		3H^2=&\kappa ^2\frac{1}{2}\dot{\phi}^2+\kappa ^2 V(\phi)+\kappa ^2 \rho_m+\frac{1}{2}(\alpha+(1- 3\omega )\beta)\rho ,\label{18}\\
		2\dot{H}+3H^2&=-\kappa^2 \frac{1}{2}\dot{\phi}^2+\kappa^2 V(\phi)-\frac{1}{2}[\alpha \omega+(1+5\omega )\beta]\rho.\label{19}
	\end{align}
	Eq. \eqref{18} can be rewritten as,
	\begin{align}
		1=& \frac{1}{2}\frac{\kappa ^2 \dot{\phi}^2}{3H^2}+\frac{\kappa ^2 V(\phi)}{3H^2}+ \frac{\kappa ^2  \rho_m}{3H^2}+\frac{1}{2}(\alpha+(1-3\omega )\beta)\frac{\rho}{3H^2}.\label{20}
	\end{align}
	
	Instead of solving the field equations analytically, which is quite cumbersome because of its non-linearity, we will perform the dynamical system analysis by considering some dimensionless variables to represent the field equations. Thereby, we shall analyze different evolutionary phases of the Universe.
	
	\section{The Dynamical system analysis} \label{sec3}
	We shall adopt the dynamical system approach to investigate the evolutionary behavior of the Universe by  introducing appropriate dimensionless variables. It may also provide significant information on dark energy, dark matter, and the early Universe. Now, corresponding to the Friedmann equation \eqref{18} of $f(R,L_m,T)$ gravity, the dimensionless variables are introduced as,
    
	\begin{align}
		x_1^2=\frac{1}{2}\frac{\kappa ^2 \dot{\phi}^2}{3H^2},\quad 
		x_2^2=\frac{\kappa ^2 V(\phi)}{3H^2},\quad x_3^2= \frac{\kappa ^2 \rho_m}{3H^2},\quad
		x_4^2=\frac{1}{2}(\alpha+(1-3\omega )\beta)\frac{\rho}{3H^2}.\label{24}
	\end{align}
	
	With respect to dimensionless variables, Eq. \eqref{20} can be expressed as
	\begin{align}\label{25}
		1=x_1^2+x_2^2+x_3^2+x_4^2.
	\end{align}

	The effective energy density for the non-relativistic matter sector, $\Omega _{m}=\frac{\kappa ^2 \rho _{m}}{3H^2}$; the energy density for the geometric dark energy sector, which is associated with the effective quantity derived from the generalized Friedmann equations of $f (R,L_m,T)$ gravity, $\Omega _{de}=\frac{1}{6H^2}(\alpha+(1-3\omega)\beta) \rho $; and the energy density for the scalar component, $\Omega _{\phi}=\frac{\kappa ^2 \rho _{\phi}}{3H^2}$. Now the expressions of $\Omega_m$, $\Omega_{\phi}$ and $\Omega_{de}$, and their relation in terms of $x_1$, $x_2$, $x_3$ and $x_4$ are obtained from the first Friedmann equation \eqref{18} as
	\begin{align}  \label{Omega}   
		1=\Omega_{m}+\Omega_{\phi}+\Omega_{de},
	\end{align}
	where $\Omega _{m}=x_3^2$, $\Omega _{\phi}=x_1^2+x_2^2$ and $\Omega _{de}=x_4^2$.
	Further, we can obtain
	\begin{align}
		\frac{\dot{H}}{H^2}=\frac{3}{2}\left[-1-\left(\frac{\alpha \omega+(1+5\omega)\beta}{\alpha+(1-3\omega)\beta}\right)x_4^2-x_1^2+x_2^2\right].\label{30}
	\end{align}

	Using Eq. \eqref{30}, the deceleration parameter ($q$) and effective equation of state parameter ($\omega_{eff}$) with respect to the dimensionless variables can be expressed as,
	\begin{align}
		q&=-1-\frac{\dot{H}}{H^2}\nonumber\\
		&=\frac{1}{2}+\frac{3}{2}\left[\left(\frac{\alpha \omega +(1+5\omega  )\beta}{\alpha+(1-3\omega  )\beta}\right)x_4^2+x_1^2-x_2^2\right],\label{31}\\
		\omega_{eff}&=-1-\frac{2\dot{H}}{3H^2}\nonumber\\
		&=\left(\frac{\alpha \omega +(1+5\omega  )\beta}{\alpha+(1-3\omega  )\beta}\right)x_4^2+x_1^2-x_2^2.\label{32}
	\end{align}
	
	Using  Eq. \eqref{11} and Eq. \eqref{12}, the equation of state parameter corresponding to the scalar field can be expressed as 
	\begin{align}\label{37}
		\omega _{\phi}&=\frac{p_{\phi}}{\rho _{\phi}}=\frac{x_1^2-x_2^2}{x_1^2+x_2^2}.
	\end{align}
	
	The following autonomous dynamical system is obtained by taking the derivative of dimensionless variables with respect to e-folding number $N=\ln{a}$. 
	\begin{align}
			\frac{dx_1}{dN}&=\frac{3x_1}{2}\left[-1+\sqrt{\frac{2}{3}}\frac{\lambda}{\kappa}\frac{(1-\omega _{\phi})}{(1+\omega _{\phi})}x_1-x_2^2+x_1^2
		\left(\frac{\alpha \omega +(1+5\omega )\beta}{\alpha+(1-3\omega  )\beta}\right)x_4^2\right],\label{38}\\
		\frac{dx_2}{dN}&=\frac{3x_2}{2}\left[1-\sqrt{\frac{2}{3}}\frac{\lambda}{\kappa} x_1-x_2^2+x_1^2 \left(\frac{\alpha \omega +(1+5\omega )\beta}{\alpha+(1-3\omega  )\beta}\right)x_4^2\right],\label{39}\\
		\frac{dx_3}{dN}&=\frac{3x_3}{2}\left[\left(\frac{\alpha \omega +(1+5\omega )\beta}{\alpha+(1-3\omega  )\beta}\right)x_4^2-x_2^2
		+x_1^2\right],\label{40}\\
		\frac{dx_4}{dN}&=\frac{3x_4}{2}\left[-(1+\omega )-x_2^2+x_1^2\left(\frac{\alpha \omega +(1+5\omega )\beta}{\alpha+(1-3\omega  )\beta}\right)x_4^2\right].\label{41}    
	\end{align}
		Equating Eqs. \eqref{38}-\eqref{41}, the critical points of the autonomous system can be obtained. Once the critical points are identified, the stability of the eigenvalues of the Jacobian matrix at each critical point is to be assessed. The Jacobian matrix captures the characteristics of linear stability and whether a critical point is saddle, unstable, or stable depends on the type of eigenvalues—positive, negative, or mixed. More information can be found in \cite{41_wiggins2003introduction}. For the autonomous system \eqref{38}-\eqref{41}, we found eight crucial points, which are represented by the symbols $A^\pm$, $B^\pm$, $C^\pm$ and $D^\pm$. In \autoref{tab:TableA}, all critical points along with their respective existence conditions are given and the corresponding eigenvalue of each critical points has been provided in \autoref{tab:TableB}. The cosmological parameters and the values of density parameters are listed in \autoref{tab:TableC}.

\begin{table}[H]
		\centering
		\renewcommand{\arraystretch}{1.2} 
		\setlength{\tabcolsep}{10pt}       
		\begin{tabular}{|c|c|c|}		\hline
			\textbf{	Critical Points }& \textbf{($\bm{x_1,x_2,x_3,x_4}$)}& \textbf{Existence}\\
			\hline
			$A^{+}$&$(0,1,0,0)$ & Always\\
			\hline
			$A^{-}$&$(0,-1,0,0)$ & Always\\
			\hline
			$B^{+}$&$\left(0,0,0,\frac{\sqrt{\omega (\alpha+(1-3\omega )\beta})}{\sqrt{\beta+\omega(\alpha+5\beta)}}\right)$ &$\beta+\omega(\alpha+5\beta)\ne0$\\
			\hline
			$B^{-}$&$\left(0,0,0,-\frac{\sqrt{\omega (\alpha+(1-3\omega )\beta})}{\sqrt{\beta+\omega(\alpha+5\beta)}}\right)$ &$\beta+\omega(\alpha+5\beta)\ne0$\\
			\hline
			$C^{+}$&$\left(\frac{(\omega_{\phi}-1)\lambda+\sqrt{6\kappa^2(1+\omega_{\phi})^2+(\omega_{\phi}-1)^2\lambda^2}}{\sqrt{6}\kappa(1+\omega_{\phi})},0,0,0\right)$ &$\omega_{\phi}\ne-1$\\
			\hline
			$C^{-}$&$\left(\frac{(\omega_{\phi}-1)\lambda-\sqrt{6\kappa^2(1+\omega_{\phi})^2+(\omega_{\phi}-1)^2\lambda^2}}{\sqrt{6}\kappa(1+\omega_{\phi})},0,0,0\right)$ &$\omega_{\phi}\ne-1$\\
			\hline
			$D^{+}$&$\left(\frac{\sqrt{\frac{3}{2}}\kappa (1+\omega_\phi)}{\lambda},\frac{\sqrt{3\kappa^2(1+\omega_{\phi})^2-2\omega_{\phi}\lambda^2}}{\sqrt{2}\lambda},0,0\right)$ &Always\\
			\hline
			$D^{-}$&$\left(\frac{\sqrt{\frac{3}{2}}\kappa (1+\omega_\phi)}{\lambda},-\frac{\sqrt{3\kappa^2(1+\omega_{\phi})^2-2\omega_{\phi}\lambda^2}}{\sqrt{2}\lambda},0,0\right)$ &Always\\
			\hline
		\end{tabular} 
	\caption{\label{tab:TableA}Critical points and the existence condition} 
	\end{table}

\begin{figure}[H]
\begin{minipage}{0.43\linewidth}
	\centering    \includegraphics[width=1\linewidth]{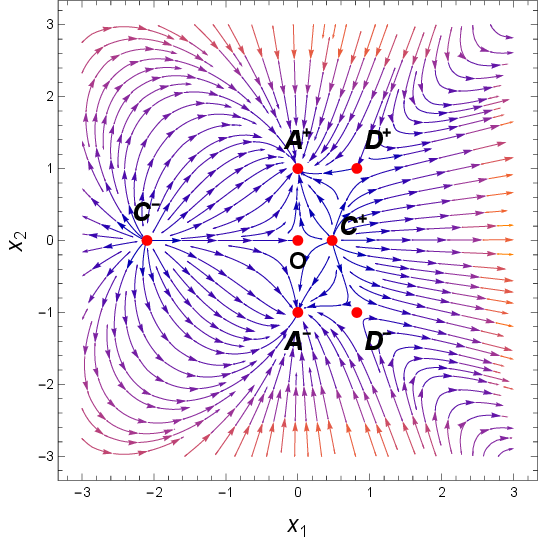}
	\caption{$x_1x_2$ plane, $\omega_\phi=-0.33$, $\omega=-1$}
	\label{fig:fig1}
	\end{minipage}
	\hfill
	\begin{minipage}{0.43\linewidth}
	\centering
	\includegraphics[width=1\linewidth]{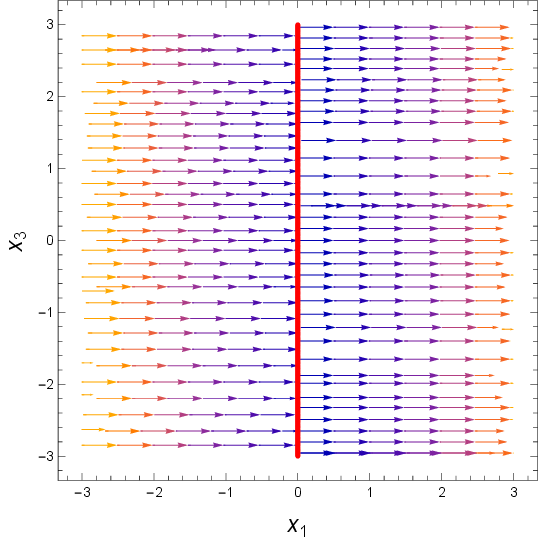}
	\caption{$x_1x_3$ plane, $\omega_\phi=-0.99$, $\omega=-1$}
	\label{fig:fig2}
	\end{minipage}
\end{figure}

\begin{figure}[H]
    
	\begin{minipage}{0.43\linewidth}
	\centering    \includegraphics[width=1\linewidth]{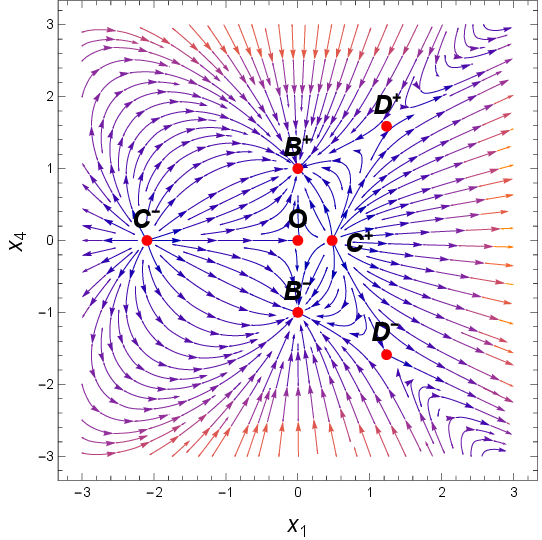}
	\caption{$x_1x_4$ plane, $\omega_\phi=-0.33$, $\omega=-1.33$}
	\label{fig:fig3}
    \end{minipage}
	\hfill
	\begin{minipage}{0.43\linewidth}
	\centering
	\includegraphics[width=1\linewidth]{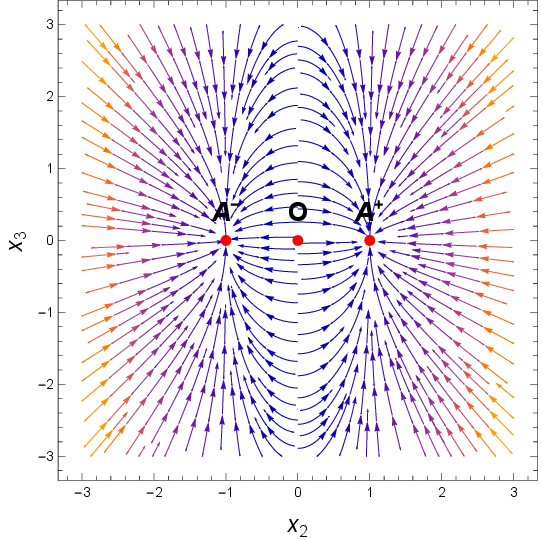}
	\caption{$x_2x_3$ plane, $\omega_\phi=-1$, $\omega=-0.33$}
	\label{fig:fig4}
    \end{minipage}
	\end{figure}

\begin{figure}[H]
	\centering
	\includegraphics[width=0.5\linewidth]{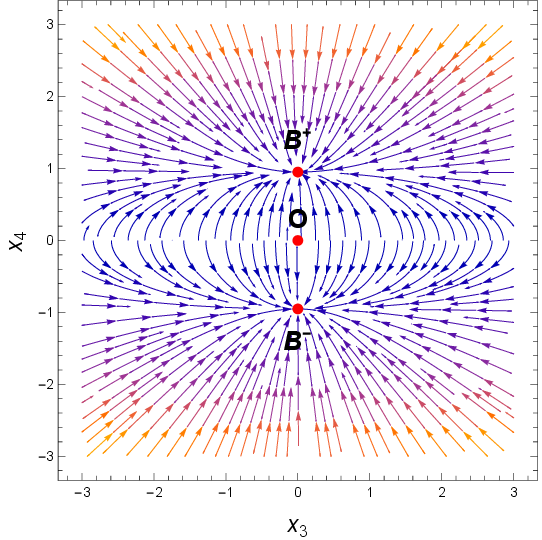}
	\caption{$x_3x_4$ plane,  $\omega_\phi=-1$, $\omega=-1.33$}
	\label{fig:fig5}
	\end{figure}

\begin{table}[H]
	\centering
	\renewcommand{\arraystretch}{1.2} 
	\setlength{\tabcolsep}{10pt}       
	\begin{tabular}{|c|c|c|c|c|c|}
		\hline
		\textbf{Points}&$\bm{\lambda_1}$  &$\bm{\lambda_2}$&$\bm{\lambda_3}$ &$\bm{\lambda_4}$\\
		\hline
		$A^{\pm}$&$-3$&$-3$&$-\frac{3}{2}$ &$-\frac{3}{2}(1+\omega )$ \\
		\hline
		$B^{\pm}$&$\frac{3}{2}(\omega -1)$ &$\frac{3}{2}\omega $ &$3\omega $ &$\frac{3}{2}(1+\omega )$ \\
		\hline
		$C^{+}$& $\frac{3}{2}+3\Psi^{+}$&$\lambda_{2C^{+}}$ &$\frac{3}{2}(1-\omega )+6\Psi^{+}$ &$3+\theta$\\
		\hline
		$C^{-}$& $\frac{3}{2}+3\Psi^{-}$&$\lambda_{2C^{-}}$ & $3+\theta$&$\frac{3}{2}(1-\omega )+\theta$\\
		\hline
		$D^{\pm}$& $-\frac{3}{2}(\omega -\omega_\phi)$&$\frac{3 \omega_\phi}{2}$ &$\frac{3}{4}(1+\omega_\phi)+\gamma$ &$\frac{3}{4}(1+\omega_\phi)-\gamma$\\
		\hline
	\end{tabular}
	\caption{\label{tab:TableB}Eigenvalues corresponding to each critical point}	
    \end{table}
    where
 \begin{align*}
 \Psi^+&=\frac{(\omega_{\phi}-1)^2\lambda^2+(\omega_{\phi}-1)\lambda\sqrt{6\kappa^2(1+\omega_{\phi})^2+(\omega_{\phi}-1)^2\lambda^2}}{6\kappa^2(1+\omega_{\phi})^2},\\
 \lambda_{2C^{+}}&=3-\frac{\lambda \left((\omega_{\phi}-1)\lambda+\sqrt{6\kappa^2(1+\omega_{\phi})^2+(\omega_{\phi}-1)^2\lambda^2}\right)}{\kappa^2(1+\omega_{\phi})^2},  \\ \Psi^-&=\frac{(\omega_{\phi}-1)^2\lambda^2-(\omega_{\phi}-1)\lambda\sqrt{6\kappa^2(1+\omega_{\phi})^2+(\omega_{\phi}-1)^2\lambda^2}}{6\kappa^2(1+\omega_{\phi})^2}, \\
 \lambda_{2C^{-}}&=3+\frac{\lambda \left((1-\omega_{\phi})\lambda+\sqrt{6\kappa^2(1+\omega_{\phi})^2+(\omega_{\phi}-1)^2\lambda^2}\right)}{\kappa^2(1+\omega_{\phi})^2}, \\
  \gamma&=\frac{3 }{4\lambda}\sqrt{24\kappa^2(1+\omega_\phi)^2+(1+(\omega_\phi-14)\omega_\phi)\lambda^2}, \\
  \theta&=\frac{3(\omega_{\phi}-1)\lambda}{(1-\omega_{\phi})\lambda+\sqrt{6\kappa^2(1+\omega_{\phi})^2+(\omega_{\phi}-1)^2\lambda^2}}. 
\end{align*}
\begin{table}[H]
	\centering
	\renewcommand{\arraystretch}{1.2} 
	\setlength{\tabcolsep}{10pt}       
	\begin{tabular}{|c|c|c|c|c|c|}
		\hline
		\textbf{Points}&$\bm{q}$&$\bm{\Omega_m}$&$\bm{\Omega_\phi}$&$\bm{\Omega_{de}} $&$\bm{\omega_{eff}}$  \\
		\hline
		$A^{\pm}$&$-1$&0&1&0&$-1$ \\
		\hline
		$B^{\pm}$&$\frac{1}{2}(1+3\omega )$&0&0&$\frac{\omega (\alpha+(1-3\omega )\beta)}{\beta+\omega (\alpha+5\beta)}$&$\omega $\\
		\hline
		$C^{+}$&$2+3\Psi^+$&0&$1+2\Psi^+$&0&$1+2\Psi^+$\\
		\hline
		$C^{-}$&$2+3\Psi^-$&0&$1+2\Psi^-$&0&$1+2\Psi^-$\\
		\hline
		$D^{\pm}$&$\frac{1}{2}(1+3\omega_{\phi})$&0&$\frac{3\kappa^2(1+\omega_\phi)^2}{\lambda^2}-2\omega_\phi$&$0$&$\omega_{\phi}$\\
		\hline
	\end{tabular}
	\caption{\label{tab:TableC}The deceleration parameter, effective equation of state parameter and density parameters}
\end{table}
	Details of each critical point are given below.
\begin{itemize}
\item \textbf{Point $A^{\pm} $:} There is always a cosmological solution that corresponds to the critical points $A^\pm$ with the coordinates $(0,\pm1 , 0,0)$. The scalar density for the points is $\Omega_\phi=1$, while the matter density and dark energy density components are $\Omega_m=0$ and $\Omega_{de}=0$, respectively. This indicates that there is no contribution of matter and dark energy to the cosmological dynamics for this specific solution and scalar field component  is dominated. The value of the deceleration parameter $q $ is $-1$ and the effective equation of state parameter is $\omega_{eff}=-1$, the negative values of these parameters indicate an accelerated expansion of the Universe. The eigenvalues are $\lambda_1=-3$, $\lambda_2=-3$, $\lambda_{3}=-\frac{3}{2}$ and $\lambda_4=-\frac{3}{2}(1+\omega)$ for these $A^\pm$ points. In quintessence region $A^\pm$ have stable behavior, while it has saddle behavior for phantom region.
\item \textbf{Point $B^{\pm} $: } The critical points exist except for $\beta+\omega (\alpha+5\beta)=0$. The coordinates of critical points $B^\pm$ are  $\left(0,0,0,\pm \frac{\sqrt{\omega(\alpha+(1-3\omega)\beta})}{\sqrt{\beta+\omega (\alpha+5\beta)}}\right)$. The matter density and scalar density components are $\Omega_m=0$ and $\Omega_{\phi}=0$. So, there is no contribution of the matter component and the scalar component at these particular points. The dark energy density is $\Omega_{de}=\frac{\omega(\alpha+(1-3\omega)\beta)}{\beta+\omega (\alpha+5\beta)}$, which means that the dark energy component  is dominated. In this case, the Universe observes accelerated expansions for $\omega<-1/3$, since the deceleration parameter is $q=\frac{1}{2}(1+3\omega)$ and the effective equation of state parameter, $\omega_{eff}=\omega$. The points $B^\pm$ have eigenvalues $\lambda_1=\frac{3}{2}(\omega-1)$, $\lambda_2=\frac{3}{2}\omega$, $\lambda_3=3\omega$ and $\lambda_4=\frac{3}{2}(1+\omega)$. These equilibrium points are stable in the phantom region, while they are saddle in quintessence region.
\item \textbf{Point $C^+$ : } For this point, the coordinates are $\left(\frac{(\omega_{\phi}-1)\lambda+\sqrt{6\kappa^2(1+\omega_{\phi})^2+(\omega_{\phi}-1)^2\lambda^2}}{\sqrt{6}\kappa(1+\omega_{\phi})},0,0,0\right)$, which always exist except for $\omega_\phi=-1$. The matter density parameter $\Omega_m$ and the dark energy density $\Omega_{de}$ vanish at these points and the only non-vanishing energy density is due to the scalar field, which is $\Omega_\phi=1+2\Psi^+$. Also, the value of the deceleration parameter is $q=2+3\Psi^+$ and the effective equation of state parameter, $\omega_{eff}=1+2\Psi^+$, so the scalar component dominates. The eigenvalues are $\lambda_1=\frac{3}{2}+3\Psi^{+}$, $\lambda_2=\lambda_{2C^+}$, $\lambda_3=\frac{3}{2}(1-\omega)+6\Psi^{+}$ and $\lambda_4=3+\theta$. All the eigenvalues for the case cannot be simultaneously negative, and in particular, the equilibrium points can be saddle or unstable.
\item \textbf{Point $C^-$ : } This point exists except at $\omega_\phi=-1$ and $\left(\frac{(\omega_{\phi}-1)\lambda-\sqrt{6\kappa^2(1+\omega_{\phi})^2+(\omega_{\phi}-1)^2\lambda^2}}{\sqrt{6}\kappa(1+\omega_{\phi})}\right. \\ \left.,0,0,0\right)$ are  its coordinates. Similarly to the point  $C^+$, in this point also there is no role for matter and the dark energy parameter. The value of the scalar density parameter, $\Omega_\phi=1+2\Psi^-$,  deceleration parameter is $q=2+3\Psi^-$ and the effective equation of state parameter, $\omega_{eff}=1+2\Psi^-$, which indicates the scalar field dominated era. At $C^-$ point, eigenvalues are $\lambda_1=\frac{3}{2}+3\Psi^{-}$, $\lambda_2=\lambda_{2C^-}$, $\lambda_3=3+\theta$ and $\lambda_4=\frac{3}{2}(1-\omega)+\theta$. All these eigenvalues are always positive. Hence, the nature of this point $C^-$ is unstable.
\item \textbf{Point $D^\pm$ : }The coordinates are $\left(\frac{\sqrt{\frac{3}{2}}\kappa (1+\omega_\phi)}{\lambda},\pm\frac{\sqrt{3\kappa^2(1+\omega_{\phi})^2-2\omega_{\phi}\lambda^2}}{\sqrt{2}\lambda},0,0\right)$ for $D^\pm$ points, which always exist for $\lambda>0$. For $D^\pm$, matter density parameter and dark energy density parameter are $\Omega_m=0$ and $\Omega_{de}=0$. Scalar density parameter $\Omega_\phi=\frac{3\kappa^2(1+\omega_\phi)^2}{\lambda^2}-2\omega_\phi$, the value of deceleration parameter $q=\frac{1}{2}(1+3\omega_\phi)$ and effective equation of state parameter $\omega_{eff}=\omega_\phi$, all these values are indicating scalar component dominating era. The value of $q$ and $\omega_{eff}$ are negative for $D^\pm$, represents  accelerating phase of the expanding Universe. The eigenvalues are $\lambda_1=-\frac{3}{2}(\omega-\omega_\phi)$, $\lambda_2=\frac{3 \omega_\phi}{2}$, $\lambda_3=\frac{3}{4}(1+\omega_\phi)+\gamma$ and $\lambda_4=\frac{3}{4}(1+\omega_\phi)-\gamma$. The nature of these eigenvalues are positive and negative both, so points $D^\pm$ behave as saddle.
	\end{itemize}
The two dimensional phase portrait of the autonomous system \eqref{38}-\eqref{41} has been shown in [Fig. \ref{fig:fig1} -- Fig. \ref{fig:fig5}]. In Fig.-\ref{fig:fig1} and Fig.-\ref{fig:fig4}, for points $A^\pm$ one can see that all trajectories move towards these points. Similar behavior has been observed for the points $B^\pm$, in Fig.- \ref{fig:fig3} and Fig.-\ref{fig:fig5}. From the converging behavior of the trajectories, the stable behavior of the critical points $A^\pm$ and $B^\pm$ has been confirmed. Also, we observe that the points $C^\pm$ and $D^\pm$ in both Fig.- \ref{fig:fig1} and Fig.-\ref{fig:fig3} have unstable and saddle behavior respectively. In Fig.-\ref{fig:fig2}, we obtain the equilibrium line.

\begin{figure}[h]
	\begin{minipage}{0.45\linewidth}
	    \centering
        \includegraphics[width=\linewidth]{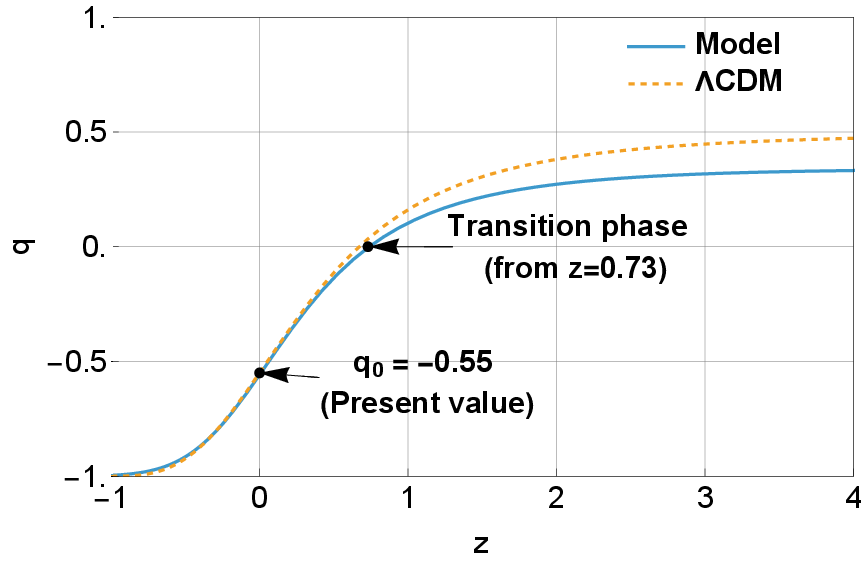}
	\caption{Deceleration parameter in redshift.}
	\label{fig:fig6}
	\end{minipage}
	\hfill
    \begin{minipage}{0.45\linewidth}
        \centering
	\includegraphics[width=\linewidth]{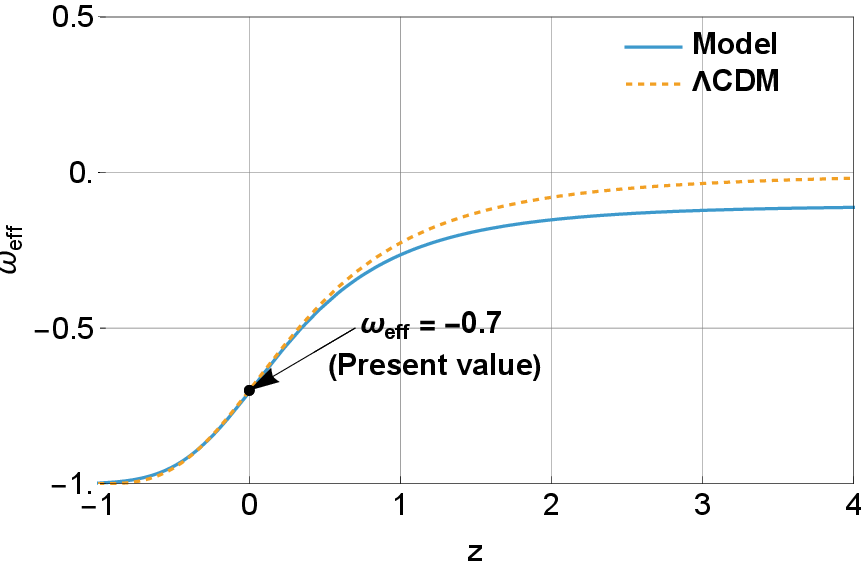}
	\caption{Dark energy equation of state parameter in redshift.}
	\label{fig:fig7}
    \end{minipage}
	\end{figure}

	The evolutionary behavior of deceleration parameter determines the accelerating or decelerating phase of the Universe. If $q$ is  positive, the Universe is in the decelerating phase and for negative $q$, it is in the accelerating phase. The deceleration parameter $q$ and  effective equation of state parameter $\omega_{eff}$ derived from the model are compared with the $\Lambda$CDM model in Fig.-\ref{fig:fig6} and Fig.-\ref{fig:fig7} respectively. In Fig.- \ref{fig:fig6}, we can observe that the Universe moves from a decelerating phase during 
    the matter-dominated epoch to an accelerating phase when dark energy starts to dominate.  The present value of $q=-0.55$ indicates that the current expansion phase of the Universe is accelerating, and this is also in line with the observational findings  \cite{52_PhysRevResearch.2.013028} and the transition is noted at the redshift, $z=0.73$. The evolutionary behavior of effective equation of state  parameter has been shown in Fig.-\ref{fig:fig7}. It can be seen that $\omega_{eff}$ is approaching $-1$ in late time and the present value is obtained to be $\omega_{eff}=-0.7$.
\begin{figure}[H]
	\begin{minipage}{0.45\linewidth}
	\centering    \includegraphics[width=\linewidth]{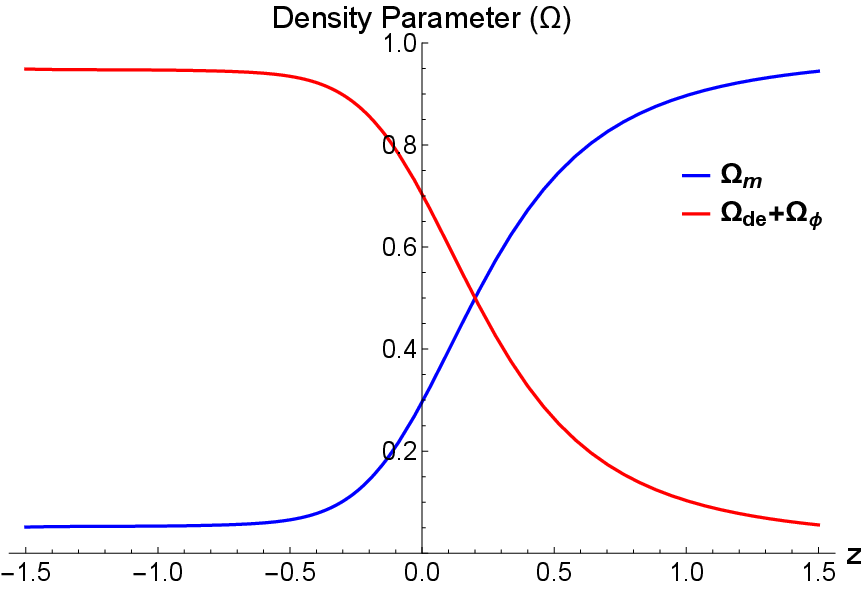}
	\caption{Density Parameter $\Omega$ in Redshift $z$}
	\label{fig:fig8}
    \end{minipage}
    \hfill
    \begin{minipage}{0.45\linewidth}
        \centering
	\includegraphics[width=\linewidth]{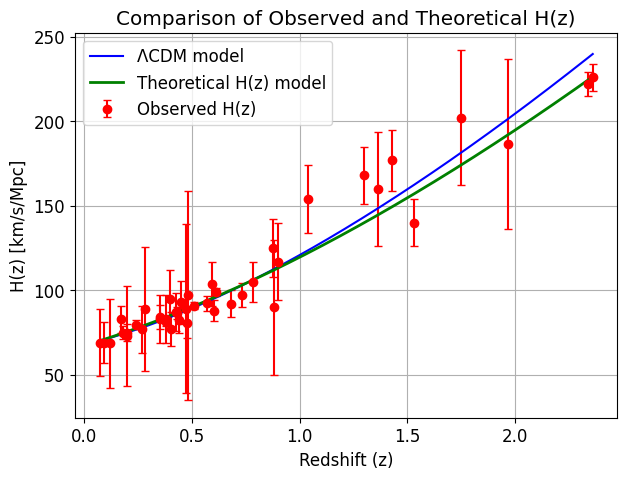} 
	\caption{Comparison with Hubble dataset}
	\label{fig:fig9}
    \end{minipage}
	\end{figure}
	Using the differential form of the Hubble parameter derived from the field equations (\ref{9}) and (\ref{10}), assuming $\alpha=1$ and $\beta=1$, we verify the accuracy of the model $f(R,L_m,T)=R+\alpha L_m+\beta T$. Compared with $H^2(z)=H_0^2(z)[\Omega_m(1+z)^3+\Omega_{de}+(1-\Omega_m-\Omega_{de})(1+z)^{2/3}]$, where $1-\Omega_m-\Omega_{de}=\Omega_\phi$, the validity of the model is verified. Fig.-\ref{fig:fig9} illustrates the $43$ data points of the Hubble data set used to examine the evolution of the Hubble parameter of the model and demonstrates its alignment with the standard $\Lambda$CDM model \cite{53_Hussain_2025}. The present value of the Hubble parameter, $H_0=67.26$, the density of matter $\Omega_m$=0.26 and the value of the combined density parameter for the dark energy component and the contribution of the scalar field  $\approx 0.74$. There is significant agreement with traditional cosmological evidence, as the proposed model aligns with the observable data.
    
	\section{Conclusion} \label{sec4}
	The evolution of the Universe at different phases has been shown using dynamical system analysis in $f(R,L_m,T)$ gravity. An additive algebraic structure has been considered in the matter-geometry coupling form of the model given by $f(R,L_m,T)=R+\alpha L_m+\beta T $ and the Friedmann equations has been transformed into a set of dimensionless variables. In addition, the energy momentum tensor is considered to be divergenceless, leading to the conclusion that the conservation equations \eqref{14}-\eqref{16} are equal to zero. An exponential potential for the scalar field has been introduced. A total of eight critical points were obtained along with their existence condition, as shown in Table--\ref{tab:TableA}. In addition, the corresponding eigenvalues of each critical point have been shown in Table--\ref{tab:TableB}. Table--\ref{tab:TableC} provides the value of the cosmological and dynamical parameters for each critical point.
	
	The stability analysis of the autonomous system reveals that the critical points describe different phases of the Universe. The points $A^+$ and $A^-$ have a stable behavior in the quintessence region, and at these points the values of deceleration parameter and the effective equation of state parameter is $-1$ which describes accelerating expansion of the Universe. For $A^\pm$, scalar density parameter $\Omega_\phi=1$, describes scalar field is dominated. Critical points $B^+$ and $B^-$ are stable in phantom region. At these points the value of deceleration parameter will always be negative, representing the late time acceleration phase of cosmic evolution and effective equation of state parameter $\omega_{eff}=\omega$, which describes the dark energy component is dominated. It is identified that the eigenvalues corresponding to points $C^\pm$ and $D^\pm$ can not be simultaneously negative. In other words, these points cannot be stable. The point $C^+$ can have both saddle and unstable behavior, while the point $C^-$ would always have unstable behavior. In both points $C^+$ and $C^-$, the scalar component is dominated and in particular, the potential part vanishes and the contribution is from the derivative of scalar field part, only.  The points $D^\pm$ will remain as saddle for all the parameter values and for these points $w_{eff}=\omega_\phi$, which indicates the domination of scalar field component. Also, we get accelerating phase of the expanding Universe for negative values of $\omega_\phi$. 
	
	The phase space diagrams for the combinations of dimensionless variables are given in Figs. \ref{fig:fig1}-\ref{fig:fig5}. In Fig.-\ref{fig:fig1} and Fig.-\ref{fig:fig4},  stable behavior of the points $A^\pm$ is shown, while Fig.-\ref{fig:fig3} and Fig.-\ref{fig:fig5} shows the converging trajectories and hence showing the stable behavior for the critical point $B^\pm$. The behavior of points $C^\pm$ and $D^\pm$ is unstable and saddle respectively, as observed in Fig.-\ref{fig:fig1} and Fig.-\ref{fig:fig3}. In Fig.-\ref{fig:fig6} an Fig-\ref{fig:fig7}, deceleration parameter $(q)$ vs redshift $z$ and effective equation of state parameter ($\omega_{eff}$) vs $z$, respectively are plotted and also in these plots are aligned with $\Lambda$CDM model. From Fig.-\ref{fig:fig6}, the present value of deceleration parameter is $q=-0.55$ confirms the accelerating phase of the Universe. The data comparison of the theoretical model with the $43$ Hubble dataset, demonstrating that it has evolved similarly to the $\Lambda$CDM model [Fig.-\ref{fig:fig9}]. Finally we conclude that using dynamical system analysis approach in $f(R,L_m,T)$ gravity, the evolutionary behavior of the Universe can be assessed and specifically the late time acceleration of the Universe can be realized.
	\subsection*{Funding}
    DS acknowledges SHODH (Scheme of Developing High Quality Research) (Ref No.2024016420), Department of Education, Government of Gujarat for providing financial support. BM thanks SPU for providing financial support to visit DoMSPU under their PM-USHA grant; during the visit the work was conceptualized.
	\subsection*{Conflict of interest}
The authors of this work declare that they have no conflict of interest.
	\small
	\bibliographystyle{unsrt}
	\bibliography{apssamp}
	
\end{document}